\documentclass[a4paper,aps,pre,preprint,floatfix,superscriptaddress,nofootinbib,showpacs]{revtex4}

\usepackage[dvips]{epsfig}
\usepackage[dvips]{graphicx}
\usepackage{latexsym,verbatim}
\usepackage{color}

\newcommand{\av}[1]{\langle {#1} \rangle}

\newcommand{\avk}{\langle k\rangle}

\begin{document}

\title{Voter models on weighted networks}

\author{Andrea Baronchelli}

\affiliation{Departament de F\'\i sica i Enginyeria Nuclear,
  Universitat Polit\`ecnica de Catalunya, Campus Nord B4, 08034
  Barcelona, Spain}

\author{Claudio Castellano}

\affiliation{Istituto dei Sistemi Complessi (ISC-CNR), Via dei Taurini 19,
I-00185 Roma, Italy}

\affiliation{Dipartimento di Fisica, ``Sapienza''
Universit\`a di Roma, P.le A. Moro 2, I-00185 Roma, Italy}

\author{Romualdo Pastor-Satorras}

\affiliation{Departament de F\'\i sica i Enginyeria Nuclear,
  Universitat Polit\`ecnica
 de Catalunya, Campus Nord B4, 08034
  Barcelona, Spain}

\date{\today}

\begin{abstract}
  We study the dynamics of the voter and Moran processes running on
  top of complex network substrates where each edge has a weight
  depending on the degree of the nodes it connects. For each
  elementary dynamical step the first node is chosen at random and the
  second is selected with probability proportional to the weight of
  the connecting edge.  We present a heterogeneous mean-field approach
  allowing to identify conservation laws and to calculate exit
  probabilities along with consensus times. In the specific case
  when the weight is given by the product of nodes' degree raised to a
  power $\theta$, we derive a rich phase-diagram, with the
  consensus time exhibiting various scaling laws depending on $\theta$
  and on the exponent of the degree distribution $\gamma$. Numerical
  simulations give very good agreement for small values of
  $|\theta|$. An additional analytical treatment (heterogeneous pair
  approximation) improves the agreement with numerics, but the
  theoretical understanding of the behavior in the limit of large
  $|\theta|$ remains an open challenge.
\end{abstract}


\maketitle

\section{Introduction}
\label{sec:introduction}

Many technological, biological and social networks are
intrinsically weighted. Each link has associated an
additional variable, called weight, which gauges the intensity or
traffic of that connection, and that can exhibit widely varying
fluctuations \cite{Barrat:2004b,Barthelemy:2005,barrat07:_archit}. The
presence of weights is extremely relevant in some scenarios
(e.g. in the case of transport in a network in which
weights measure bandwidth or capacity), and it must therefore be 
taken into account explicitly. Some results have already been produced in this
direction, dealing, among other problems, with diffusive processes
\cite{wu07:_walks,dynam_in_weigh_networ}, epidemic spreading
\cite{Yan05,colizza07:_invas_thres}, general equilibrium and
non-equilibrium phase transitions
\cite{PhysRevLett.95.098701,karsai:036116}, or glassy dynamics
\cite{PhysRevE.80.020102}.  
Here we present a detailed investigation of the ordering dynamics of
voter-like models on weighted networks \cite{Castellano09}.
 
The voter model~\cite{Clifford73,holley1975etw} and the Moran
process~\cite{moran1962spe} are simple examples of
ordering dynamics, which allow to understand how natural systems with an
initial disordered configuration are able to achieve order via local
pairwise interactions.  Both models are described in terms of a
collection of individuals, each endowed with a binary variable $s_i$,
taking the values $\pm 1$. The elementary step 
consists in randomly choosing a first individual and then (again
randomly) one of her nearest neighbors. In the voter model the first
individual will \textit{copy} the state of her neighbor. In the Moran
process, on the other hand, she will \textit{transmit} her own state
to the neighboring node, which will adopt it.  In both cases, starting
from a disordered initial state, the iteration of the elementary step
leads to the growth of correlated domains and, in finite systems, to
an absorbing uniform state in which all individuals share the same
state (the so-called consensus).  In a social science context, the
voter model represents thus the simplest model of opinion
formation in a population, in which individuals can change their
opinion as a function of the state of their nearest neighbors
\cite{Castellano09}. In the same way, in a biological context, the
Moran process represents the elementary example of two species competing
(through reproduction and neutral selection) for the same environment
\cite{nowak2006ed}.

The voter and Moran processes are equivalent on regular lattices and
on the complete graph, but if the pattern of connections 
is given by a complex (unweighted) topology they behave differently, since the
order in which interacting individuals are selected becomes
relevant~\cite{Suchecki05,castellano05:_effec}. Moreover, the time to reach consensus 
scales with the system size in different ways depending on the degree distribution of the
network~\cite{Sood05,Antal06,Sood08}. Considering a weighted
topological substrate adds, as we will see, a richer and
more complex phenomenology. Additionally, the case of weighted
networks allows to model very natural settings.
 In the context of social sciences, for example, weights can reflect the
obvious fact that the opinion of a given individual can be more easily
influenced by a close friend rather than by a casual acquaintance. On
the other hand, in an evolutionary scenario, weights allow to gauge the
effects of heterogeneous replacement rates in different species.

On weighted networks, at each time step a vertex $i$ is
selected randomly with uniform probability; then one nearest neighbor
of $i$, namely $j$, is chosen with a probability proportional to the
weight $w_{ij} \ge 0$ of the edge joining $i$ and $j$. That is, the
probability of choosing the neighbor $j$ is
\begin{equation}
  P_{ij} = \frac{w_{ij}}{\sum_r w_{ir}}.
  \label{eq:18}
\end{equation}
Vertices $i$ and $j$ are then updated according to the rules of the
respective models.  With this definition, the models considered
represent the natural extension for ordering dynamics on weighted
networks (and in particular of the voter model) of the generalized
Moran process proposed in Refs.~\cite{lieberman2005edg,baxter:258701},
in which dynamics was defined as a function of a set of arbitrary
interaction probabilities $P_{ij}$. In our case, however, the fact
that these interaction probabilities arise from the normalized weights
arriving at a vertex imposes some restrictions to the possible values
of $P_{ij}$ and yields therefore different outcomes and
interpretations. Also, it is worth noting three recent publications
\cite{schneider-mizell09,Yang09,Lin10}, dealing with related, but not
identical, models.

Adopting the heterogeneous mean-field (HMF) approximation
\cite{dorogovtsev07:_critic_phenom,barratbook} we will assume that the
weight between vertices $i$ and $j$ depends only on the degrees at the
edges endpoints, namely $k_i$ and $k_j$, and therefore we can write
$w_{i j} \equiv g(k_i, k_j) a_{ij}$, where $a_{ij}$ is the adjacency
matrix and $g(k,k')$ is a positive definite, symmetric function.  The
application of HMF theory and the backwards Fokker-Planck
formalism~\cite{Sood05,Antal06,Sood08}, allows us to derive analytical
expressions in degree uncorrelated networks for the main relevant
quantities (namely, exit probability and consensus time
\cite{Castellano09}) in a more transparent way than in
Refs.~\cite{lieberman2005edg,baxter:258701}.  In order to allow for
closed mathematical solutions of the models, we will further specify
the function $g$ to be given by the product of two independent
functions $g(k,k')=g_s(k) g_s(k')$, an assumption motivated by
empirical observations in real weighted networks~\cite{Barrat:2004b}.
Specializing both models to the case of networks with power-law
distributed degrees and edge weights given by multiplicative powers
the endpoint degrees, $g_s(k)=k^\theta$, a very rich phase-diagram is
obtained, with several different scaling regions of the consensus time
with the network size $N$. A numerical check of the analytical
predictions reveals a good agreement in some regions of the parameters
space and noticeable discrepancies in others.  In order to gain
insights into the observed numerical disagreement, we apply an
improved mean-field approach, the heterogeneous pair
approximation~\cite{Pugliese09}, which turns out to provide better
agreement with numerics for small $\theta$ but is still not able to
solve the problems for large $\theta$.  The qualitatively different
nature of the dynamics for large $\theta$ is briefly discussed and its
understanding identified as an intriguing challenge for future work.

\section{Heterogeneous mean-field theory} 

In this Section, we perform a theoretical analysis of the voter and
Moran processes on weighted networks within a HMF
approximation \cite{dorogovtsev07:_critic_phenom}, extending the
Fokker-Planck formalism developed for the unweighted case in
Refs.~\cite{Sood05,Antal06,Sood08}. Let us consider the models defined
by the interaction probability Eq.~(\ref{eq:18}), where the network
weights take the form
\begin{equation}
  \label{eq:19}
  w_{ij} = g(k_i,k_j) a_{ij}.
\end{equation}
The simplest way to extend the Fokker-Planck approach to weighted
networks is to follow the annealed weighted network approximation
introduced in Ref.~\cite{dynam_in_weigh_networ}.  The key point
consists in considering the degree coarse-grained
interaction probability $P_w(k \to k')$, defined as the probability
that a vertex of degree $k$ interacts with a nearest neighbor vertex 
of degree $k'$. In unweighted networks, this probability simply takes 
the form of the conditional probability $P(k'|k)$ that a vertex of degree $k$ 
is connected to a vertex of degree $k'$ \cite{alexei}.  In networks with 
weights given by Eq.~(\ref{eq:19}), the interaction probability of the voter/Moran
dynamics, Eq.~(\ref{eq:18}), can be coarse-grained by performing an
appropriate degree average, to yield \cite{dynam_in_weigh_networ}
\begin{equation}
  P_w(k \to k') = \frac{g(k, k')  P(k'|k)}{ \sum_q g(k,q)
    P(q|k)}. 
  \label{eq:10}
\end{equation}

The relevant function defining voter and Moran processes is the
probability $\Pi(k; s)$ that a spin $s$ at a vertex of degree $k$
flips its value to $-s$ in a microscopic time step
~\cite{Sood05,Antal06,Sood08}. This function can be expressed, within
the annealed weighted network approximation, in terms of the density
$x_k$ of $+1$ spins in vertices of degree $k$, taking the form
\begin{eqnarray}
  \Pi_V(k; +1) &=& P(k) x_k \sum_{k'} P_w(k \to k')
  (1-x_{k'}), \label{eq:11} \\
  \Pi_V(k; -1) &=&  P(k) (1-x_k) \sum_{k'} P_w(k \to k')
  x_{k'}, \label{eq:12}           
\end{eqnarray}
for the voter model. The origin of these probabilities is easy to
understand \cite{Sood05,Antal06,Sood08}. For example,
Eq.~(\ref{eq:11}) gives the probability of flipping a vertex of degree
$k$ in the state $+1$ as the product of the probability $P(k)$ of
choosing a vertex of degree $k$, times the probability $x_k$ that the
vertex is in the state $+1$, times the probability $k$ chooses to
interact with a neighbor vertex $k'$, which is in state $-1$ with
probability $1-x_{k'}$, averaged over all possible neighbor degrees
$k'$. Analogously, the flipping probabilities for the Moran process
can be expressed as
\begin{eqnarray}
  \Pi_M(k; +1) &=& \sum_{k'} P(k') (1-x_{k'}) P_w(k' \to k)
  x_k, \label{eq:13} \\ 
  \Pi_M(k; -1) &=&  \sum_{k'} P(k') x_{k'} P_w(k' \to k)
  (1-x_k).\label{eq:14} 
\end{eqnarray}

Let us now present separately the mean-field analysis for the two
models under consideration.

\subsection{Voter model}

\subsubsection{Rate equation, conservation laws and exit probability}

Let us consider the time evolution of the density $x_k$, which is
determined in terms of a rate equation. Following
\cite{Sood05,Antal06,Sood08,castellano05:_effec,dynam_in_weigh_networ},
this rate equation is shown to take the form
\begin{eqnarray}
  \dot{x}_k(t) &=&  \sum_{k'} P_w(k \to k')  x_{k'}(t) - x_k(t)
  \nonumber \\
  &=&  \sum_{k'}  \frac{g(k, k')  P(k'|k)}{ \sum_q
    g(k,q)  P(q|k)} x_{k'}(t) - x_k(t) ,
  \label{eq:15}
\end{eqnarray}
where in the last expression we have used Eq.~(\ref{eq:10}). The
complete expression Eq.~(\ref{eq:15}), valid for any correlation and
weight patterns, is quite difficult to deal with.  In order to obtain
closed analytical expressions, we assume that the underlying network
is degree uncorrelated, namely, $P(k'|k) = k' P(k') / \avk$
\cite{mendesbook}, and moreover, that the weights are simple
multiplicative functions of the edges' end points, that is, $g(k, k')
= g_s(k) g_s(k')$.  In this way, Eq.~(\ref{eq:15}) becomes
\begin{equation}
  \dot{x}_k(t) =  \omega_V(t) - x_k(t),
  \label{eq:1}
\end{equation}
where we have defined
\begin{equation}
  \omega_V(t)= \sum_{k'}   \frac{k' g_s(k')P(k')}{\av{ k
      g_s(k)}} x_{k'}(t) , 
      \label{e:omega_v}
\end{equation} 
and $\langle f(k) \rangle \equiv \sum_k P(k) f(k)$.

It is easy to see that the total density of $+1$ spins,
$x =\sum_k P(k) x_k$ is not a conserved quantity,
$\dot{x} = - x +\omega_V$. The
quantity $\omega_V$, however, is conserved, $\dot{\omega}_V(t)=0$, as
we can see by inserting Eq.~(\ref{eq:1}) into the time derivative of
$\omega_V(t)$.  
Finally, the steady state condition of
Eq.~(\ref{eq:1}), $\dot{x}_k =0$, implies $x_k = \omega_V$

As for the usual voter model~\cite{Sood05,Sood08} the conservation law
allows the immediate determination of the exit (or ``fixation")
probability $E$, i.e. the probability that the final state
corresponds to all spins in the state $+1$. 
In the final state with all $+1$ spins we have
$\omega_V=1$, while $\omega_V=0$ is the other possible final state
(all $-1$ spins).  Conservation of $\omega_V$ implies then
$\omega_V = E \cdot 1+ [1-E] \cdot 0$, hence
\begin{equation}
E = \omega_V.
\end{equation}
Starting from an homogeneous initial condition, with a density $x$ of
randomly chosen vertices in the state $+1$, we obtain, since $\omega_V = x$,
$E_h(x) = x$
as in the standard voter model \cite{Castellano09}. On the other hand,
with initial conditions consistent of a single $+1$ spin in a vertex
of degree $k$, we have $E_1(k) = k g_s(k)/[N \av{k g_s(k)}]$.

\subsubsection{Consensus time}

The backward Fokker-Planck formalism \cite{Gardinerbook} can be applied to
obtain expressions for the consensus time $T_N(x)$, as a function of
the initial density $x$ of $+1$ spins and the system size $N$.
However, following Refs.~\cite{Sood05,Antal06,Sood08,castellano05:_effec}, 
it is simpler to apply a one-step calculation 
and use the recursion relation \cite{Sood05,Antal06,Sood08}
\begin{equation}
  T_N(\left\{ x_k \right\})= \sum_s \sum_k \Pi_V(k; s)[T_N(x_k
  -s\Delta_k) + \Delta t]
 + Q(\left\{ x_k \right\}) [T_N(\left\{ x_k
  \right\}) + \Delta t],
  \label{eq:6}
\end{equation}
where $Q=1-\sum_s \sum_k \Pi_V(k; s)$ is the probability than
no spin flip takes place, $\Delta t=1/N$ and $\Delta_k = 1/[NP(k)]$
is the change in
$x_k$ when a spin flips in a vertex of degree $k$.  Rearranging the
terms in Eq.~(\ref{eq:6}), and expanding to second order in
$\Delta_k$, we obtain
\begin{equation}
  \frac{1}{2 N} \sum_k \frac{x_k+ \omega_V - 2 x_k
    \omega_V}{P(k)} \frac{\partial^2 T_N}{\partial 
    x_k^2} 
  + \sum_k (x_k - \omega_V)  \frac{\partial T_N}{\partial
    x_k} = -1.
  \label{eq:5}
\end{equation}
Eq.~(\ref{eq:5}) is simplified by observing that the ordering dynamics
of the voter model is separated in two well distinct temporal
regimes~\cite{Sood05,Vazquez08}. Over a short time the different
densities $x_k$ all converge from their initial value to the common
value at the steady state $x_k=\omega_V$.  For infinite-size systems,
this state survives forever.  For finite size $N$, the system enters
instead a different regime where the dynamics of densities $x_k$ is
enslaved by the fluctuations of $\omega_V$, which performs a slow
diffusion until it hits the absorbing values $0$ or $1$. The consensus
time is dominated by this second regime. This allows to apply the
steady state condition which cancels the drift term
in Eq.~(\ref{eq:5}).  Taking as relevant quantity the conserved
weighted magnetization $\omega_V$, we obtain \cite{Sood05,Sood08}
\begin{equation}
  \frac{1}{N}  \omega_V (1-\omega_V) 
  \frac{\av{k^2 g_s(k)^2}}{\av{k g_s(k) }^2} \frac{\partial^2
    T_N}{\partial 
    \omega_V^2}=-1. 
\end{equation}
The integration of this equation leads to 
\begin{displaymath}
  T_N(\omega_V) = -N
  \frac{\av{k g_s(k) }^2}{\av{k^2 g_s(k)^2}} \left[ \omega_V
    \ln 
    \omega_V + (1-\omega_V) \ln (1-\omega_V) \right].
\end{displaymath}
Thus, the ordering time starting from homogeneous initial conditions,
$x_k = \omega_V=1/2$ is
\begin{equation}
  T_N(x=1/2) = N (\ln 2) \frac{\av{k g_s(k) }^2}{\av{k^2 g_s(k)^2}}.
\end{equation}

\subsection{Moran process}

\subsubsection{Rate equation, conservation laws and exit probability}

The derivation of the rate equation for the density $x_k$ in the Moran
process follows the same steps as in the voter model, taking the form
\begin{eqnarray}
  \dot{x}_k(t) &=& \frac{1}{P(k)} \sum_{k'} P(k') P_w(k' \to k) [ x_k(t)
  - x_{k'}(t) ] \nonumber \\
  &=& k  \sum_{k'} \frac{P(k'|k)}{k'} \frac{g(k'k)}
{\sum_q  g(k',q) P(q|k')} [x_k(t)-x_{k'}(t)], \nonumber
\end{eqnarray}
where in the last step we used the degree detailed balance
condition $k P(k) P(k'|k) = k' P(k') P(k|k')$ \cite{marian1}.
Assuming again a degree uncorrelated network, and multiplicative
weights, we are led to 
\begin{equation}
    \dot{x}_k(t) = \frac{k g_s(k)}{\av{k g_s(k)}} [ x_k(t) - x(t)]. 
\end{equation}
Again, the total density of $+1$ spins, $x=\sum_k P(k) x_k$ is not
conserved, while instead the quantity 
\begin{equation}
  \omega_M =  \frac{1}{\av{[k g_s(k)]^{-1}}} 
\sum_k \frac{P(k)}{k g_s(k)} x_k
  \label{eq:2}
\end{equation}
is conserved, $\dot{\omega}_M =0$.  Finally, from the steady state
condition, $\dot{x}_k =0$, we obtain $x_k = x$.
From the conservation of $\omega_M$ the exit probability is immediately
derived as
\begin{equation}
  E = \omega_M.
\end{equation}
Homogeneous initial conditions lead again to $E_h(x)=x$, while a
single $+1$ spin in a vertex of degree $k$ leads to 
\begin{equation}
  E_1(k) = \frac{1}{k g_s(k)} \frac{1}{N\av{[k g_s(k)]^{-1}}}.
\end{equation}

It is interesting to note that in the conserved quantity of the voter model, $\omega_V$, each density $x_k$ is weighted with the product $kg_s(k)$ (Eq.~(\ref{e:omega_v})), while in the correspondent $\omega_M$ for the Moran process the weight is precisely the inverse, namely $(kg_s(k))^{-1}$ (Eq.~(\ref{eq:2})). As noted in the case of unweighted networks \cite{Sood08}, intuitively this can be ascribed to the fact that in the voter model it is the first selected node that may change its state, while in the Moran process it is the second one. Thus in the voter model small-degree nodes change their state more often than high-degree nodes, and weighting them with the probability of being chosen ($kg_s(k)$) compensates this disparity leading to the conserved quantity $\omega_V$. Vice versa in the Moran process low-degree nodes change their state less often than high degree nodes, and the inverse weighting balances this difference \cite{Sood08}. 

\subsubsection{Consensus time}

Following the same steps presented for the voter model, and performing
the appropriate expansion to second order in $\Delta_k$, we obtain the
equation
\begin{eqnarray}
  &&  \sum_k  \frac{k g_s(k)}{\av{k g_s(k)}}  (x_k - x)  \frac{\partial
    T_N}{\partial 
    x_k} \nonumber \\
&&+ \frac{1}{2 N} \sum_k  \frac{k g_s(k)}{\av{k g_s(k)}} 
   \frac{x_k + x - 2 x_k x}{P(k)} \frac{\partial^2 T_N}{\partial 
    x_k^2}= -1. \nonumber
\end{eqnarray}
The steady state condition, $x_k = x$, leads to the cancellation
of the drift term. The diffusion term is simplified by changing
variables with the conserved quantity $\omega_M$, 
leading to the equation
\begin{equation}
   \frac{1}{N} \frac {1}{\av{[k g_s(k)]^{-1}}}
\frac{\omega_M (1- \omega_M) }{\av{k g_s(k)}} 
   \frac{\partial^2 T_N}{\partial \omega_M^2} =-1,
\end{equation}
where we have used the fact that, in the steady state, $x=\omega_M$.
The solution of equation for the consensus time leads now to
\begin{equation}
T_N(\omega_M) 
 =  -N \av{k g_s(k)} \av{[k g_s(k)]^{-1}}
\left[ \omega_M \ln(\omega_M) + (1-\omega_M) \ln(1-\omega_M)
   \right]. 
\end{equation}
Thus, starting from homogeneous initial conditions, $x_k=x=\omega_M = 1/2$,
we have
\begin{equation}
  T_N(x=1/2) = N (\ln 2) \av{k g_s(k)} \av{[k g_s(k)]^{-1}}. 
\end{equation}


\section{Networks with power-law degree distribution and weight strengths}
\label{sec:networks-with-power}

The actual behavior of the exit probability and  the consensus time
depends, in view of the previous calculations, on the topological
properties of the network under consideration, as well as on the
strength of the weights, as given by the function $g_s(k)$. In this
Section we consider explicitly these dependencies for the particular
case of networks with a power-law degree distribution form, $P(k) \sim
k^{-\gamma}$, and a weight strength scaling also as a power of the
degree, $g_s(k) = k^{\theta}$. This last selection is reasonable in
view of the weight patterns empirically observed in real networks
\cite{Barrat:2004b}. Let us focus on the consensus time with
homogeneous ($x=0.5$) initial conditions for the two models
considered.

In the case of the voter model, the ordering time with homogeneous
initial conditions and weights scaling as a
power of $k$ takes the form 
\begin{equation}
  T_N(1/2) = N \ln(2) \frac{\av{k^{1+\theta} }^2}{\av{k^{2+2\theta}}}. 
  \label{FullT_N}
\end{equation}
From this expression, we can obtain different scalings with the
network size $N$, depending on the characteristic exponents $\gamma$
and $\theta$; we consider only $\gamma >2$. Using the fact that
$\av{k^a} \sim \mathrm{const.}$ for $a<\gamma-1$ and $\av{k^a} \sim
k_c^{a+1-\gamma}$ for $a>\gamma-1$, where $k_c$ is the upper network
cutoff, and, in view of the comparison with numerical results for the
Uncorrelated Configuration Model \cite{ucmmodel}, considering the
scaling $k_c \sim N^{1/2}$ for $\gamma<3$ and $k_c \sim
N^{1/(\gamma-1)}$ for $\gamma>3$ \cite{mariancutofss}, we obtain the
following scaling for consensus the time:
\begin{equation}
  \label{eq:21}
  T_N(1/2) \sim \left\{
  \begin{array}{cl}
     N^{(3-\gamma)/2}, \qquad  &  \theta>\gamma-2, \;\gamma<3 \\
     \mathrm{const.} \qquad &  \theta>\gamma-2, \;\gamma>3 \\
     N^{(\gamma-2\theta-1)/2} \qquad & \gamma-2 > \theta >
     (\gamma-3)/2, \;\gamma<3\\
     N^{2(\gamma-\theta-2)/(\gamma-1)} \qquad & \gamma-2 > \theta >
     (\gamma-3)/2, \;\gamma>3\\
      N  \qquad  & \theta < (\gamma-3)/2
  \end{array}
  \right. .
\end{equation}
\begin{figure}[t]
  \begin{center}
    \includegraphics*[width=0.40\textwidth]{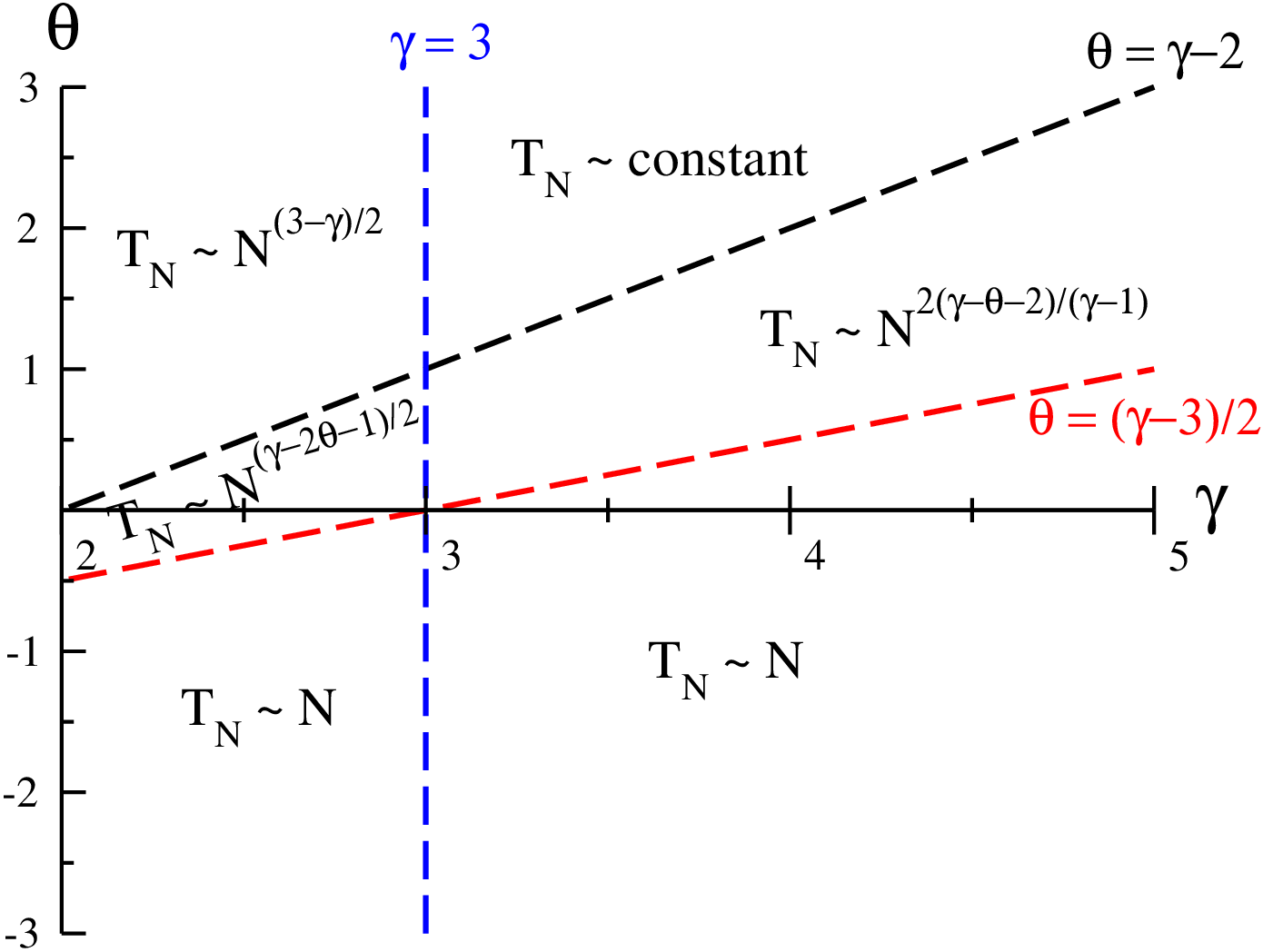} \hspace*{1cm}
    \includegraphics*[width=0.40\textwidth]{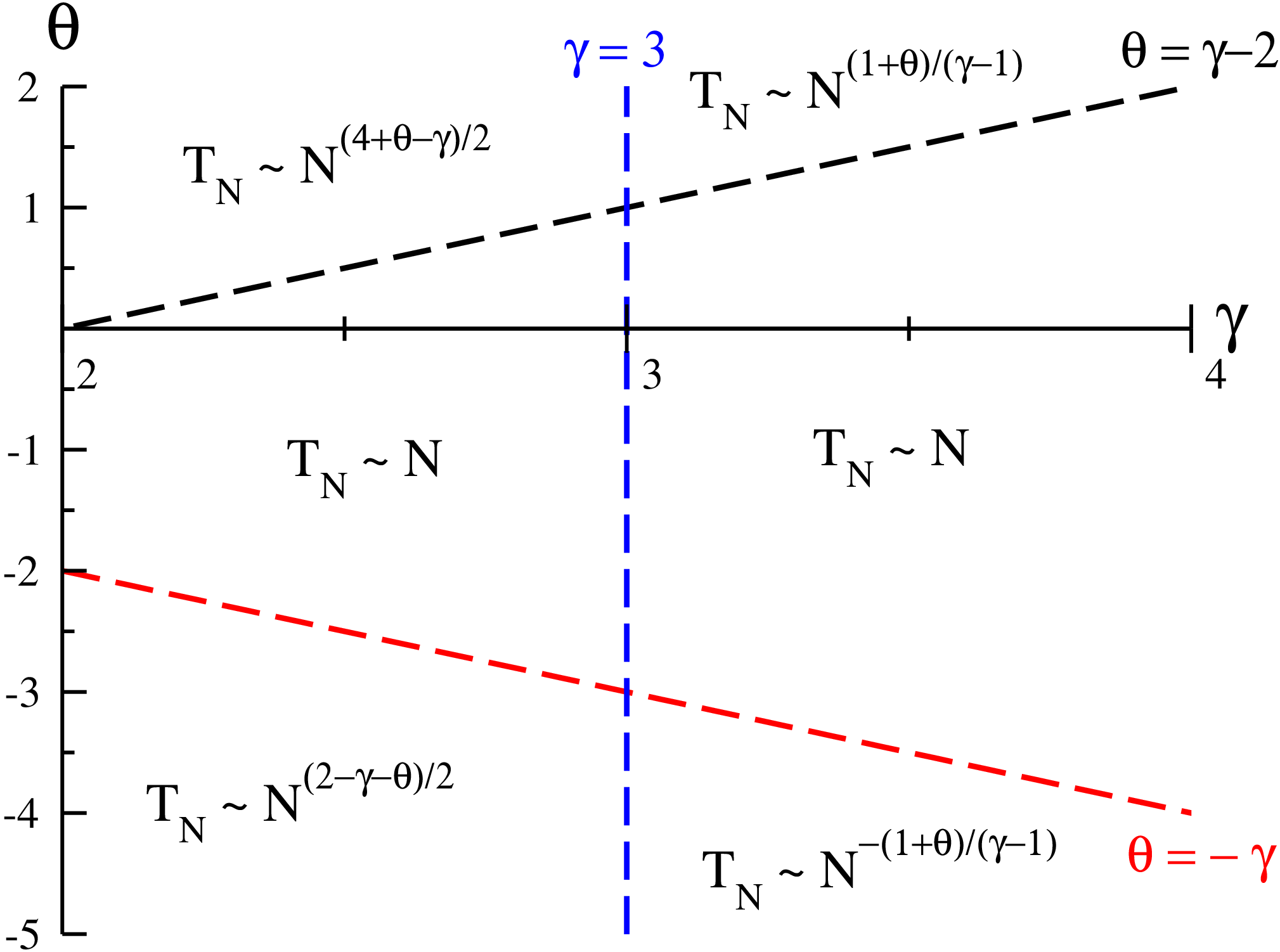}
  \end{center}
  \caption{(Color online) Phase diagram of the voter model (left) and
    Moran process (right) on weighted
    scale-free networks.}
  \label{fig:phasediagram}
\end{figure}
In Fig.~\ref{fig:phasediagram} (left) we represent graphically the different
scalings of the consensus time $T_N$ in the $(\theta, \gamma)$ space.

For the Moran process, the ordering time scales with the network size
through the expression
\begin{equation}
  T_N(1/2) =  N \ln(2) \av{k^{1+\theta} } \av{k^{-1 -\theta}}
  \label{eq:16}
\end{equation}
For $\gamma >2$, the different possible scalings are as
follows:
\begin{equation}
  \label{eq:22}
  T_N(1/2) \sim \left\{
  \begin{array}{cl}
     N^{(4+\theta-\gamma)/2}, \qquad  &  \theta>\gamma-2, \;\gamma<3 \\
     N^{(1+\theta)/(\gamma-1)}, \qquad &  \theta>\gamma-2, \;\gamma>3 \\
      N, \qquad & -\gamma <\theta<\gamma-2\\
     N^{-(1+\theta)/(\gamma-1)}, \qquad & \theta<-\gamma, \;\gamma>3\\
      N^{(2-\theta-\gamma)/2}, \qquad & \theta<-\gamma, \;\gamma<3
  \end{array}
  \right. .
\end{equation}
Fig.~\ref{fig:phasediagram} (right) depicts the different regimes
associated to the behavior of $T_N$ in the $(\theta, \gamma)$
space.

Some comments are now in order.  First we notice that all relevant
quantities are in fact function of the combination
$k^{\theta+1}$. This implies that for $\theta = -1$ both voter and
Moran dynamics are predicted to give the same results at the
mean-field level, independently of the degree distribution. In fact,
$\theta=-1$ implies that both interacting vertices are extracted
completely at random (independently of their degree) so that the
asymmetry distinguishing the voter model from the Moran process
vanishes. For other values of $\theta$, on the other hand, the effect
of weights appears to be completely different for the two dynamics.
For the voter model, positive values of $\theta$ tend to reduce the
consensus time, while $\theta<0$ leads to increased $T_N$, i.e. the
dynamics becomes slower. In any case the consensus time is at most
proportional to the system size $N$: The dynamics is always relatively
fast.  Interestingly, the HMF analysis predicts the presence of a
region ($\theta>\gamma-2$ and $\gamma>3$) for which the consensus time
is constant, i.e. the dynamics undergoes an instantaneous ordering
process, in contrast with what happens in other regions, in which
ordered regions of opposite states can coexist for very long times,
reaching consensus only in finite systems and through a large
stochastic fluctuation \cite{Castellano03}. As it will be shown below,
this is true only on annealed networks in which the quenched disorder 
imposed by the actual connections in the network is not considered. Numerical simulations
performed on quenched graphs give different results.

For the Moran process, on the other hand, $T \sim N$ represents a
lower bound for the scaling of the consensus time: The dynamics is
always rather slow, with an exponent larger than $1$ for all $(\gamma,
\theta)$.  Remarkably, the scaling of $T_N$ turns out to depend
symmetrically on $|\theta+1|$: A large positive or a large negative
value of $\theta+1$ are equally effective in slowing down the ordering
process.

\section{Comparison with numerical simulations}
\label{sec:numerical-simulations}

\subsection{Algorithms}

In order to check the analytical predictions for
the voter model and Moran process, we have performed numerical
simulations of both models on uncorrelated networks generated using
the Uncorrelated Configuration Model (UCM) \cite{ucmmodel}. The
networks have a degree exponent $\gamma$, a minimum degree
$k_m = 4$ and a maximum degree smaller than or equal to $\sqrt{N}$,
preventing the generation of correlations for $\gamma<3$
\cite{mariancutofss}. A weight strength $g_s(k) = k^\theta$ is imposed
by selecting a nearest neighbor $j$ of a vertex $i$ with probability
\begin{equation}
  P_{ij} = \frac{k_j^\theta}{\sum_{v \in \mathcal{V}(i)} k_v^\theta},
\end{equation}
where $\mathcal{V}(i)$ is the set of nearest neighbors of $i$. 

Moreover, since HMF equations describe in an exact way dynamics taking
place on annealed networks \cite{dynam_in_weigh_networ}, we have
simulated the voter model and the Moran process also on such
structures, in order to provide a benchmark of our analytical
results. In annealed networks, in fact, all links are rewired
at each microscopic time step, so that no dynamical correlation can build up, and
the absence of correlations assumed by mean-field approaches is actually implemented.
In weighted networks, the probability that a vertex of degree
$k$ interacts with a vertex of degree $k'$ is given by
\begin{equation}
   P_w(k \to k') =\frac{k' g_s(k') P(k')}{\av{k g_s(k)}} =
   \frac{k'^{1+\theta} P(k')}{\av{k^{1+\theta}}}, 
\end{equation}
where in the last equality we have assumed again that $g_s(k) =
k^\theta$. An annealed weighted network is thus implemented by
choosing as neighbor of any given vertex another vertex of degree $k$,
randomly chosen in the network with probability proportional to
$k^{1+\theta}$ \cite{dynam_in_weigh_networ}. 
In a quenched network, on the other hand, the neighbors of the first node 
are of course fixed and the choice is restricted to them.



\subsection{Exit probability}
\label{sec:exit-probability}

\begin{figure}[t]
  \begin{center}
    \includegraphics*[width=0.45\textwidth]{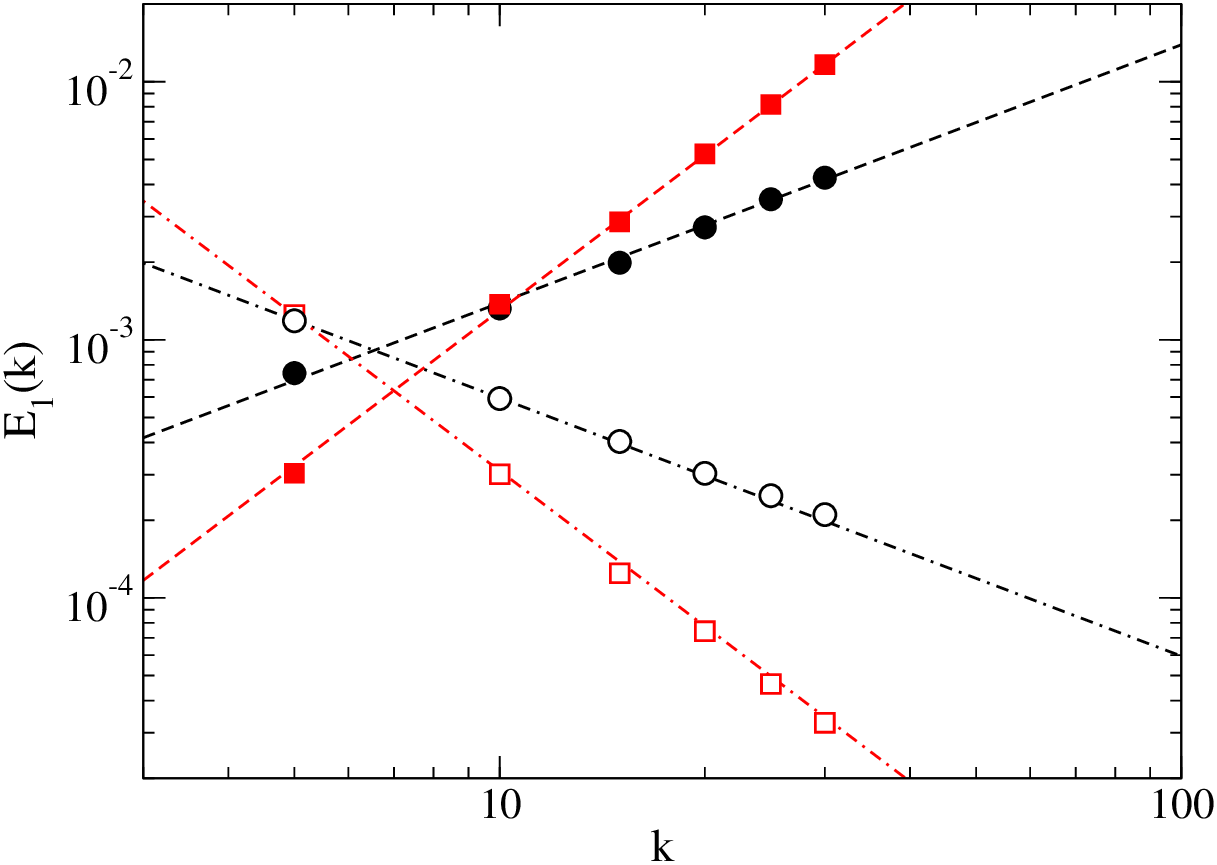}
\end{center}
\caption{(Color online) Exit probability $E_1(k)$ starting from a
  single $+1$ spin in a vertex of degree $k$, for the voter model
  (full symbols) and the Moran process (empty symbols).  Dashed lines
  represent the expected theoretical scaling with $k$, circles refer
  to the case $\theta=0$ and squares to $\theta=1$. Data from quenched
  networks of size $N=10^3$ with $\gamma=2.5$ (voter model) and
  $\gamma=2.2$ (Moran process).}
  \label{fig:exitprob}
\end{figure}

While for homogeneous initial conditions both the
voter model and Moran process lead to an exit probability equal to the
standard voter model, i.e. $E(x) = x$, invasion initial conditions
starting from a single $+1$ spin in a vertex of degree $k$ lead to
exit probabilities that depend explicitly on the initial degree
considered. In particular, we find
\begin{equation}
  E_1^\mathrm{voter}(k) \sim k^{1+\theta}, \qquad
  E_1^\mathrm{Moran}(k) \sim k^{-(1+\theta)}.
  \label{eq:17}
\end{equation}
While for the voter model  a single $+1$ vertex has better chances
to invade the system if it starts from a high degree vertex, for the
Moran process the situation is precisely the opposite, a single $+1$
spin being favored when initially located in the vertices of smallest
degree. This kind of behavior is actually to be expected from the very
definition of the models, and has been already reported in unweighted
networks \cite{Sood08}. In fact, a high degree is beneficial in the voter model since it corresponds to a larger probability of being chosen as a partner by a neighbor in search for an opinion to copy, while in the Moran process having many neighbors implies a larger probability to be invaded by the opinion at one of them. In Fig.~\ref{fig:exitprob} we plot the values
of the exit probability $E_1(k)$ computed from numerical simulations.
The results fit quite nicely the mean-field predictions in
Eq.~(\ref{eq:17}): The larger the weight intensity, the stronger the
impact of high and low degree vertices in the voter and Moran
processes, respectively.

\subsection{Consensus time}
\label{sec:consensus-time-sim}

\begin{figure*}[t]
  \begin{center}
    \includegraphics*[width=7.5cm]{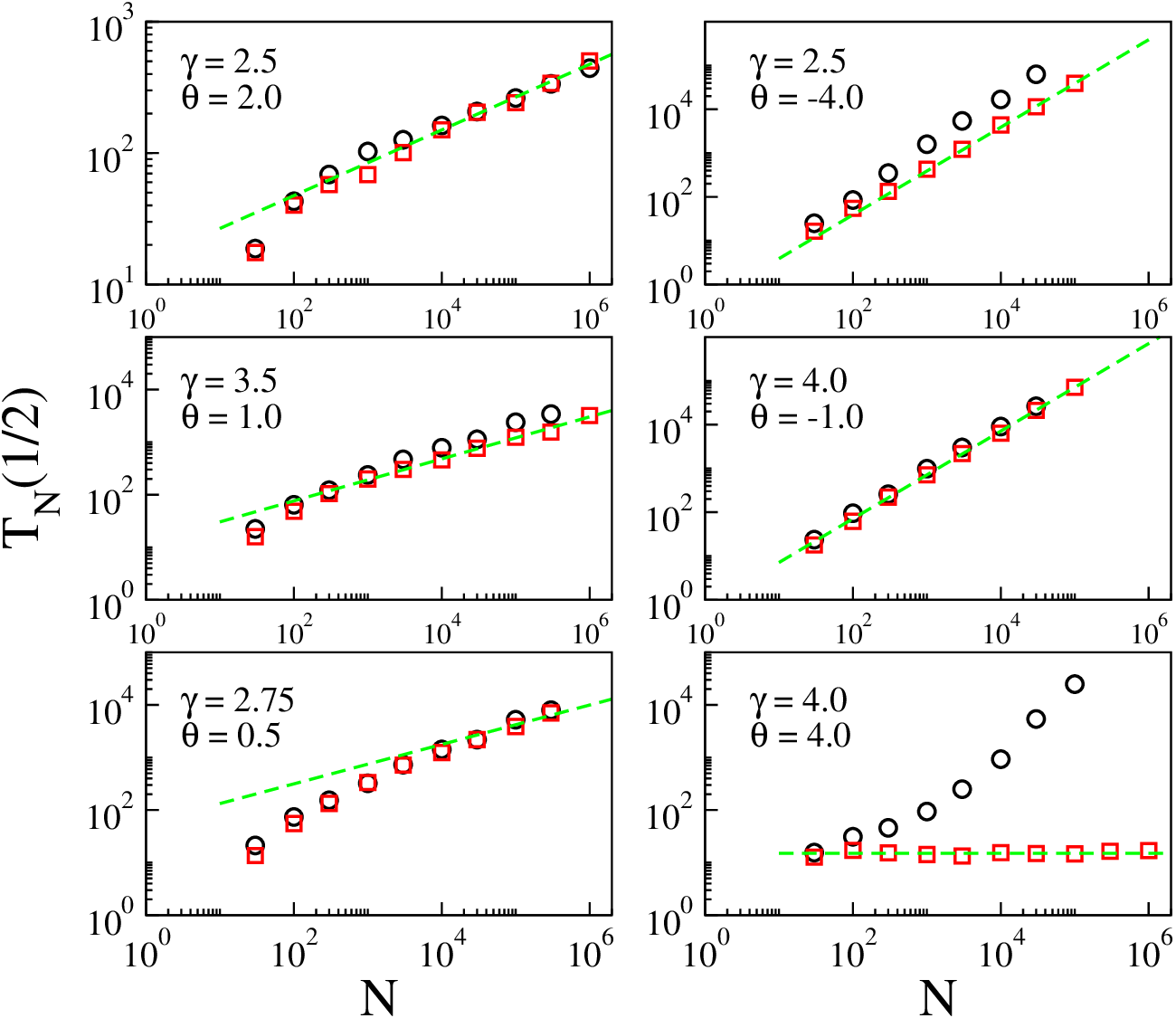} \hspace*{0.5cm}
    \includegraphics*[width=7.5cm]{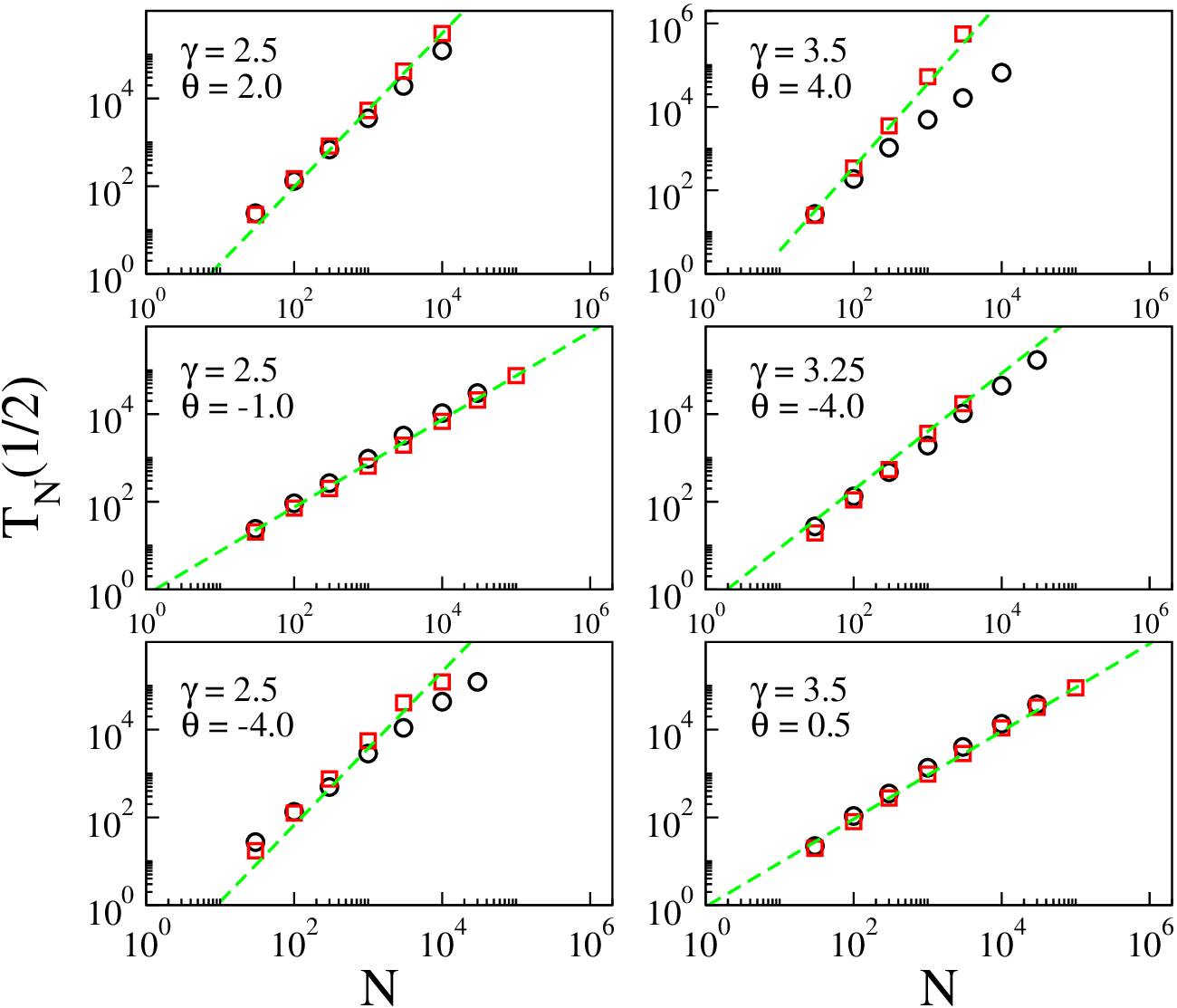}
\end{center}
\caption{(Color online) Scaling with $N$ for the voter model (left)
  and Moran process (right) on scale-free weighted networks in
  different regions of the corresponding phase diagrams,
  Fig.~\ref{fig:phasediagram}.  Squares
  represent data from simulations run on annealed networks, while
  circles concern quenched graphs. Dashed lines represent the
  theoretical scaling predicted by HMF theory.}
  \label{fig:scalingVM}
\end{figure*}

\begin{figure}[t]
  \begin{center}
    \includegraphics*[width=8cm]{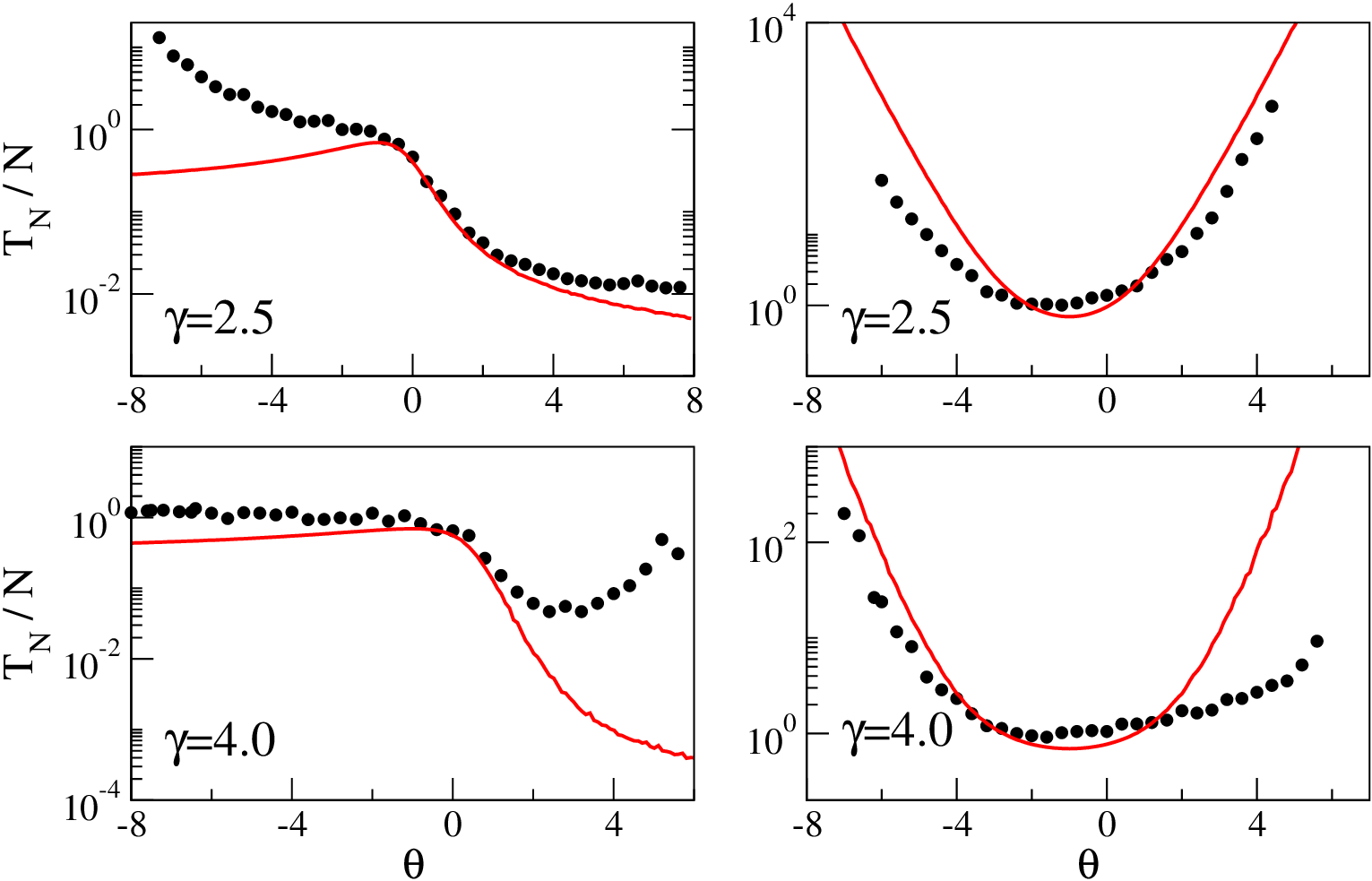}
\end{center}
\caption{(Color online) Consensus time for the voter model (left) and
  Moran process (right) in slices of the phase diagrams at two fixed
  values of $\gamma$ and varying $\theta$, compared with the
  corresponding HMF predictions, Eqs.~(\ref{FullT_N})
  and~(\ref{eq:16}), respectively (full lines). Results from
  simulations performed on quenched networks of size $N=3 \times
  10^3$.}
  \label{fig:slices}
\end{figure}

In Fig. \ref{fig:scalingVM} we check the validity of the scaling
behaviors predicted by the HMF treatment
and sketched in Fig.~\ref{fig:phasediagram}.
In this figure, we plot the
scaling of the consensus time $T_N$ as a function of $N$, for
different points in the six regions in which the respective phase
diagrams are divided, compared with the corresponding theoretical
mean-field predictions.

Fig.~\ref{fig:scalingVM} shows that, overall, the agreement between
the scaling predicted by theory and numerical data in annealed
networks is, as expected, very good. With respect to the results for
quenched networks, the agreement between HMF theory and simulations is
in general restricted to small absolute values of $\theta$, as
reported for other dynamical processes \cite{dynam_in_weigh_networ}.
In order to set better limits to the validity of the HMF
approximation, in Fig.~\ref{fig:slices} we report the numerical values
of the consensus time obtained from simulations in quenched networks
of fixed size $N=3 \times 10^3$ in slices of the phase diagrams in
Fig.~\ref{fig:phasediagram} performed at
two constant values of $\gamma$, one larger and one smaller than 3,
and varying $\theta$. These numerical values are compared with
numerical evaluations of the theoretical predictions in
Eqs.~(\ref{FullT_N}) and~(\ref{eq:16}). 
From Fig.~\ref{fig:slices} we
observe that the HMF approximation yields reasonably correct results
except for large values of $\theta$ (for $\gamma>3$) or 
large values of $-\theta$ (for $\gamma<3$). When these errors occur,
consensus time is underestimated by HMF for the voter model,
while it is overestimated for the Moran process.
At the present stage we are not able to predict a priori when
the theoretical results fail to describe the behavior of the dynamics
taking place on quenched networks, but the numerical
evidence suggests that the theory works well for values of $|\theta|$
of the order of those observed in real networks
\cite{Barrat:2004b}. 

\section{Heterogeneous pair approximation}

%
\begin{figure}[t]
  \begin{center}
    \includegraphics*[width=0.45\textwidth]{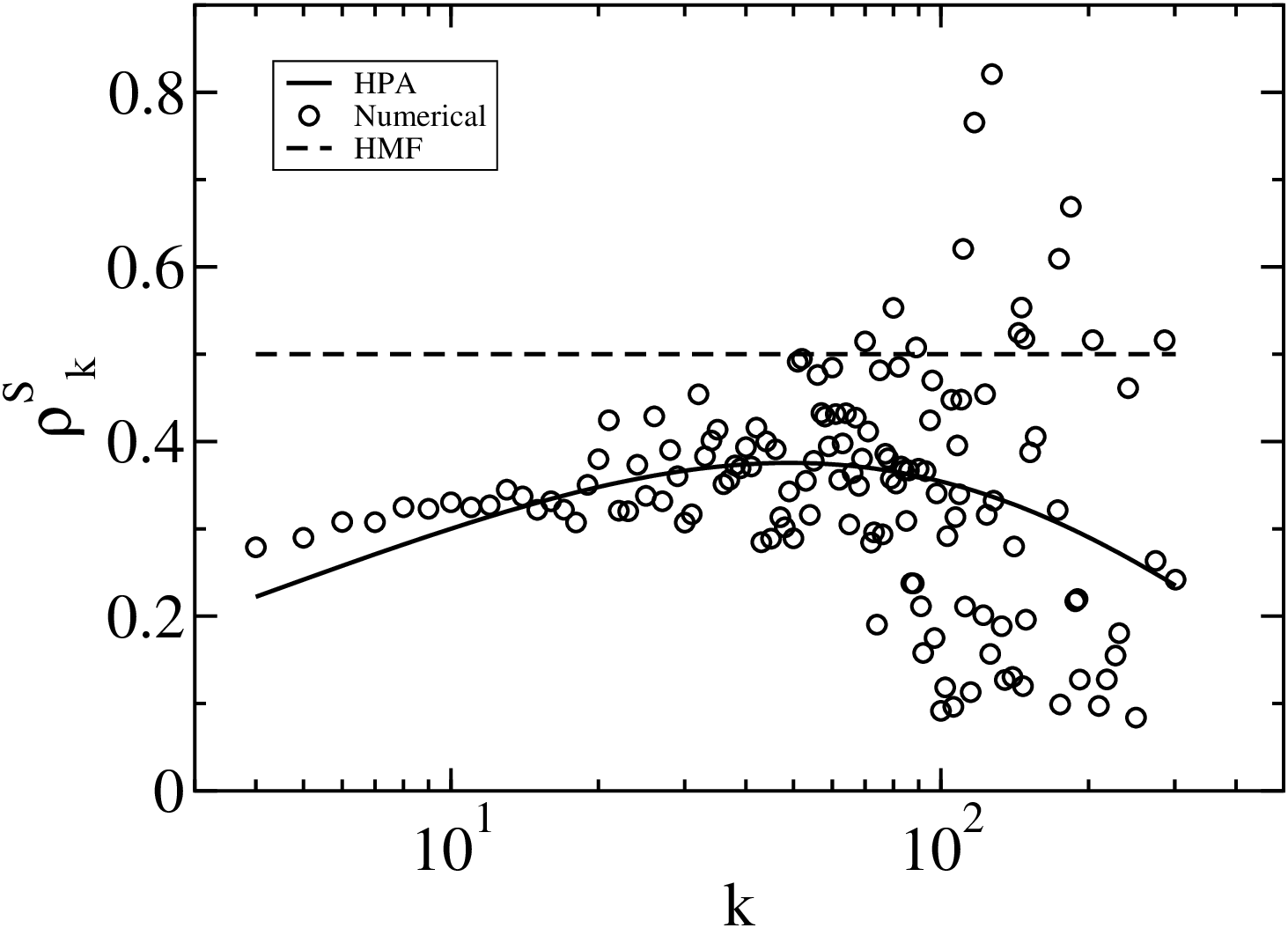}
  \end{center}
  \caption{(Color online) Probability $\rho^S_k$ that an edge
    connected to a node of degree $k$ and selected for the dynamics is
    active as a function of $k$, in the quasi steady state for
    $x=1/2$.  The solid line is the result of the heterogeneous pair
    approximation, while symbols are results of numerical simulations.
    Binning has deliberately been avoided to show the large variability 
    of numerical results for larger degrees.
    Data from voter dynamics on quenched networks of size $N=10^5$,
    with $\gamma=3.25$ and $\theta=1.5$.}
  \label{rho_k}
\end{figure}
There are several possible assumptions in the HMF treatment
which could fail when the approach breaks down.  One is the
assumption that the time to reach consensus is dominated by the
diffusive wandering of the quasi steady state, which is much larger
than the time to reach such state.
More important is however the possibility that the
very first hypothesis at the core of mean-field theory, namely that
the dynamics of the system can be fully described in terms of the
densities $x_k$, breaks down~\cite{Vazquez08,Pugliese09}.  This
assumption can be violated at several different levels.  A mild
violation occurs when the probability of a node to be in a $+1$ state
is correlated with the state of its nearest neighbors.  In order to
ascertain this possibility, it is useful to consider the quantity
$\rho_k$, defined as the probability that an edge connected to a node
of degree $k$ and selected for the dynamics is active, i.e. it
connects nodes in a different state.  Focusing on the case of the
voter model, HMF theory, which explicitly assumes the lack of
dynamical correlations between the vertices at the ends of any edge,
predicts that this quantity should be equal to
\begin{eqnarray}
  \rho_k &=& \sum_{k'} P_w(k \to k') \left[ x_k (1-x_{k'}) +
    (1-x_k) x_{k'} \right] \nonumber \\
  &=& x_k (1-\omega_V) + (1-x_k) \omega_V 
\end{eqnarray}
and hence, for initial homogeneous conditions $x_k=x=\omega_V=1/2$, we
should have in the stationary state $\rho^S_k = 2 x(1-x) =1/2$.

Fig.~\ref{rho_k} shows that, for a case where mean-field is not
accurate, this assumption is not correct in two respects: Firstly, the
value of $\rho_k$ is lower than $1/2$ (dashed line), indicating that,
in fact, correlations build up in the system. Secondly, $\rho_k$
depends on $k$, implying that those correlations depend moreover on
the degree of the nodes.

In order to take into account these degree-dependent dynamical
correlations, one needs to consider, as relevant dynamical variable,
the probability $\rho_{k,k'}$ that an edge connecting a node of degree
$k$ with another node of degree $k'$ is active, i.e. the two nodes are
in a different state.  This approach, termed heterogeneous pair
approximation (HPA), has been introduced and applied to voter models
on unweighted networks in Ref.~\cite{Pugliese09}. To determine the
equation of motion of the quantity $\rho_{k,k'}$ in the voter model,
we observe that this quantity is modified if the flipping node has
degree $k$ and one of its neighbors has degree $k'$ (or vice versa).
Let us assume that the flipping (first selected) node has degree $k$
and call $k''$ the degree of the copied (second selected) node.  It is
useful to consider separately the two cases where $k'' \ne k'$ or $k''
= k'$.

In the first case the variation $\Delta \rho_{k,k'}$ for a single
dynamical step (occurring over a time $\Delta t = 1/N$) is determined
as follows: The probability that a node in state $s$ and degree $k$
flips is given by the probability $P(k)$ that the first node selected
has degree $k$ times the probability $\sigma(s)$ that it is in state
$s$, times the probability $P_w(k \to k'')$ that the second has degree
$k''$ multiplied by the probability $\rho_{k,k''}/[2 \sigma(s)]$ that
the link connecting the two is active.  One has then to multiply this
quantity by the associated variation of the fraction of active links
between $k$ and $k'$. 
Among the $k-1$ other links of the flipping node, the number of those
connecting to a node of degree $k'$ will be $j$ distributed according
to a binomial $R(j,k-1)$ with probability of the single event equal to
$P(k'|k)$.  In their turn, only $n$ out of these $j$ links will be
active, with $n$ binomially distributed ($B(n,j)$) with single event
probability $\rho_{k,k'}/[2 \sigma(s)]$. Finally one has to multiply
by the variation of $\rho_{k,k'}$ when $n$ out of $j$ links go from
active to inactive as a consequence of the flipping of the node in
$k$.  This is given by the variation of the number of active links
$[(j-n)-n]$ divided the total number of links between nodes of degree
$k$ and $k'$, namely $N k P(k) P(k'|k)$.  One has then to sum over
$k'' \ne k'$, $s$, $j$ and $n$, obtaining
\begin{equation}
  \label{Deltarho1}
  \Delta \rho_{k,k'} =  P(k) \sum_s \sigma(s) 
  \sum_{k''\ne k'} P_w(k \to k'') \frac{\rho_{k,k''}}{2\sigma(s)} \cdot
  \sum_{j=0}^{k-1} R(j,k-1) \sum_{n=0}^j B(n,j) \frac{j-2n}{N k P(k) P(k'|k)}.
\end{equation}
By performing explicitly the summations (and using $\sum_s
1/\sigma(s)=4/(1-m^2)$, where $m=2x-1$ is the magnetization) the
formula becomes
\begin{equation}
  \frac{\Delta \rho_{k,k'}}{\Delta t} = \sum_{k''\ne k'} P_w(k \to k'') \rho_{k,k''}
  \frac{(k-1)}{k}\left(1-\frac{2}{1-m^2} \rho_{k,k'} \right).
\end{equation}
When $k''=k'$, the value of $\Delta \rho_{k,k'}$ is similar to
Eq.~(\ref{Deltarho1}) with (obviously) $P_w(k \to k')$ instead of
$P_w(k \to k'')$, no sum over $k''$, and in the numerator of the
last factor $j+1-(n+1)-(n+1)=j-2n-1$, because there are $j+1$
links to nodes of degree $k$, $n+1$ of which are active in the
initial state and inactive in the final. Summing up the two
contributions and adding the symmetric terms with $k$ and $k'$
swapped, we get

\begin{figure}[t]
  \begin{center}
    \includegraphics*[width=6.5cm]{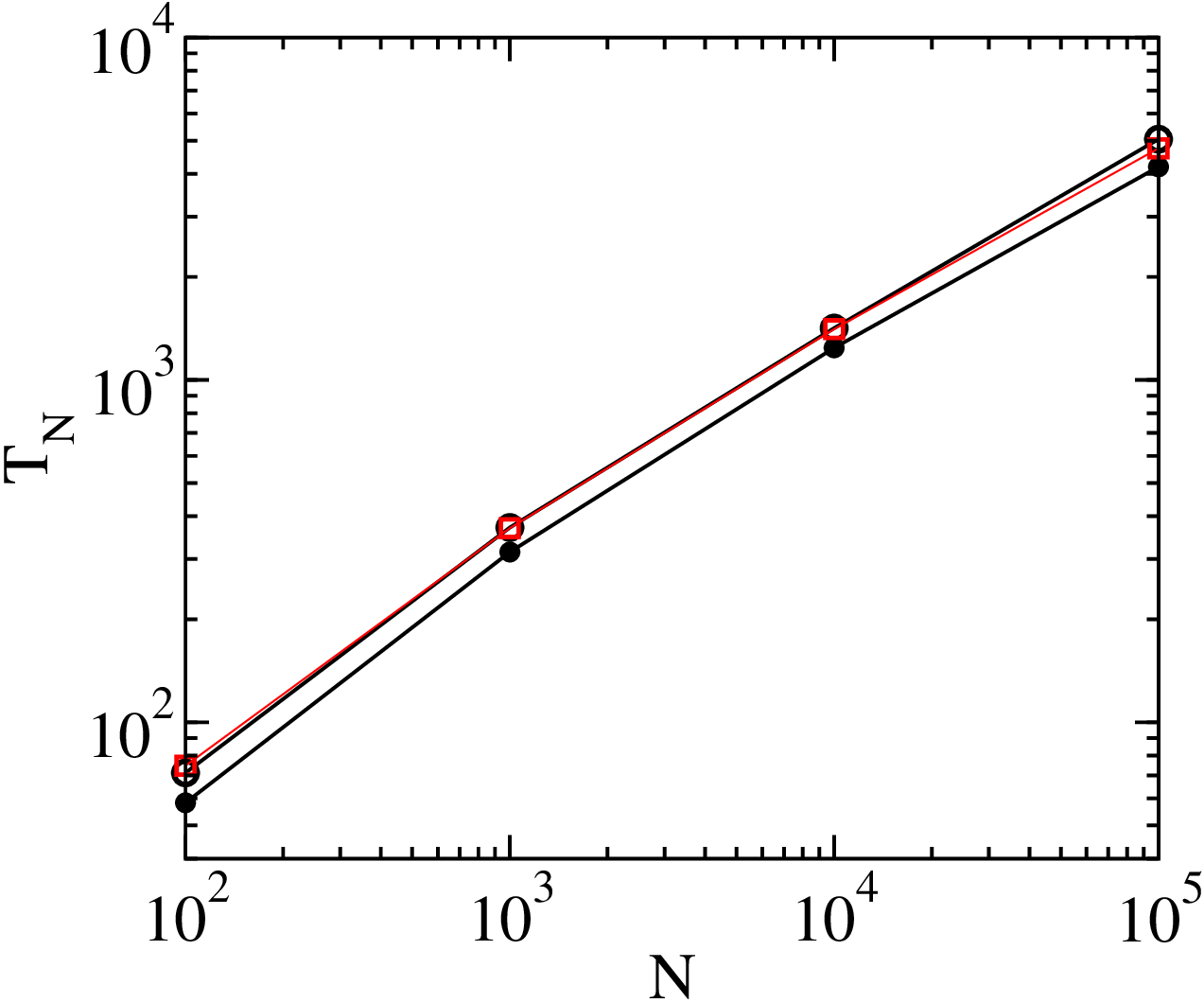} 
    \includegraphics*[width=6.5cm]{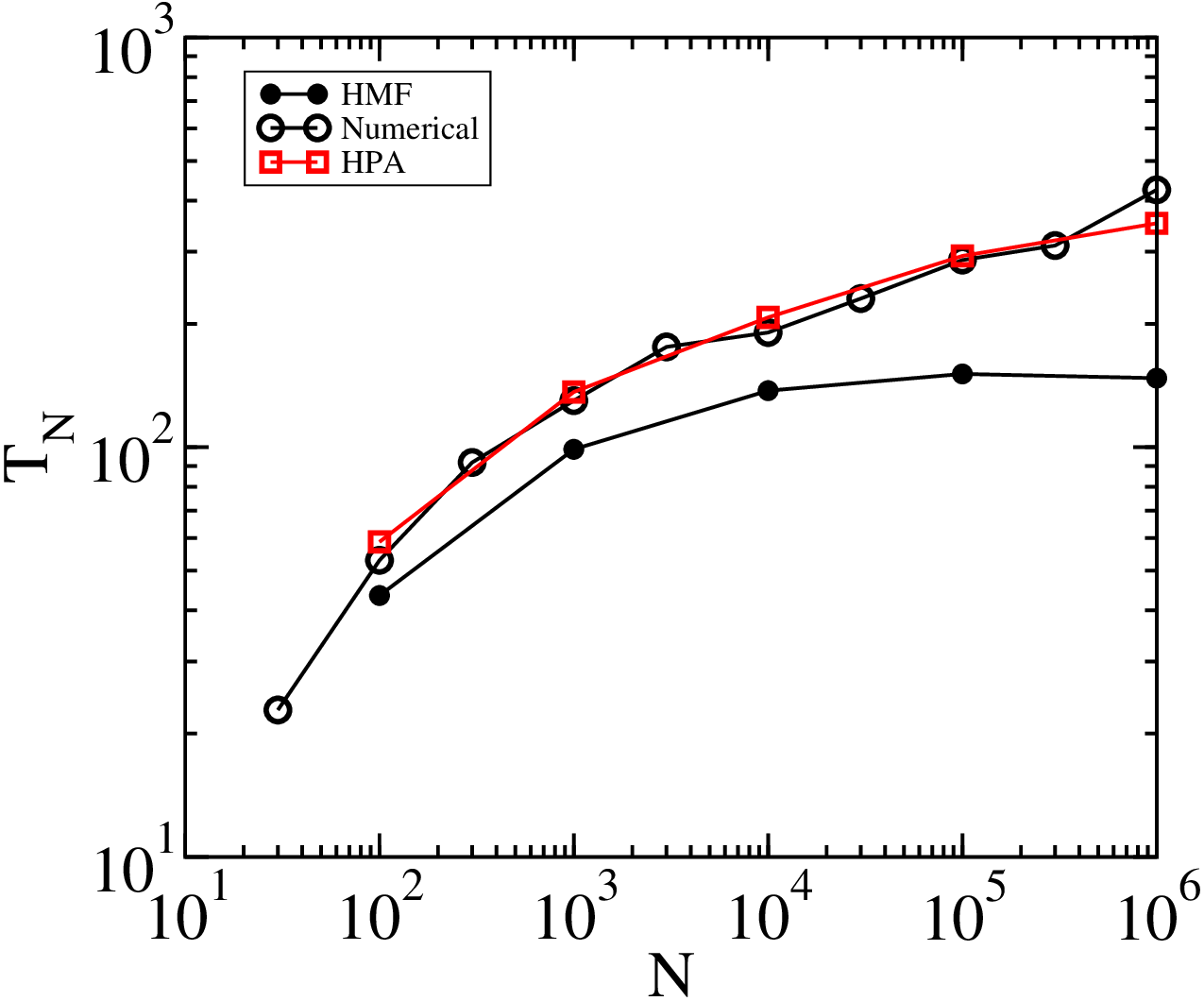}
    \includegraphics*[width=6.5cm]{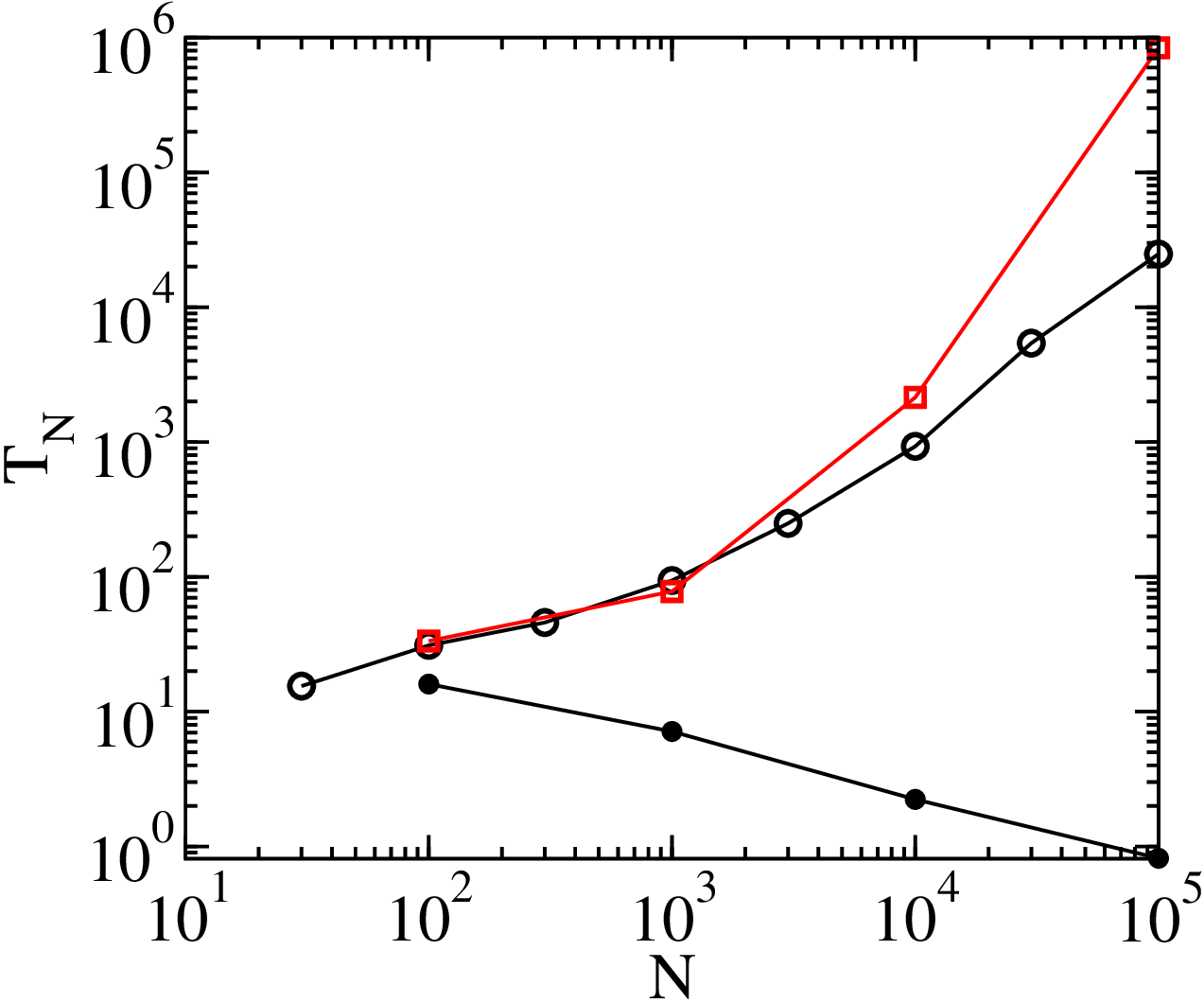}
  \end{center}
  \caption{(Color online) Comparison of the consensus time $T_N$ as a
    function of $N$ obtained in numerical simulations for voter
    dynamics on quenched networks (empty circles) and the results of
    the numerical evaluation of the HMF, Eq.~(\ref{FullT_N}) (filled
    circles), and HPA, Eq.~(\ref{eq:20}) (empty squares)
    predictions. Data correspond to networks with $\gamma=2.75$,
    $\theta=0.5$ (top), $\gamma=3.25$, $\theta=1.5$ (center), and
    $\gamma=4$, $\theta=4$ (bottom).}
  \label{T_HPA}
\end{figure}

\begin{equation}
  \frac{d\rho_{k,k'}}{dt}=\rho_k\frac{k-1}{k}+\rho_{k'}\frac{k'-1}{k'}+
  -\rho_{k,k'}\left[
    \frac{P_w(k \to k')}{P(k'|k)} \frac{1}{k}
    +\frac{P_w(k'\to k)}{P(k|k')} \frac{1}{k'}
    +\frac{2\rho_k}{1-m^2}\frac{k-1}{k}+
    \frac{2\rho_{k'}}{1-m^2}\frac{k'-1}{k'}\right].
\end{equation}
When uncorrelated networks are considered, so that
\begin{equation}
  \frac{P_w(k \to k')}{P(k'|k)} = \frac{\av{k}}{\av{k^{1+\theta}}} k'^{\theta},
\end{equation}
we are led to the final equation
\begin{equation}
  \label{drhokk'_voter}
\frac{d\rho_{k,k'}}{dt}=\rho_k\frac{k-1}{k}+\rho_{k'}\frac{k'-1}{k'}
  -\rho_{k,k'}\left[ \frac{\av{k}}{\av{k^{1+\theta}}} 
\left(
\frac{k'^{\theta}}{k}+\frac{k^{\theta}}{k'} \right)
+\frac{2\rho_k}{1-m^2}
\frac{k-1}{k}+\frac{2\rho_{k'}}{1-m^2}\frac{k'-1}{k'}\right].  
\end{equation}
where, $m$ is the magnetization and, at odds with the
case of unweighted networks, the definition of $\rho_k$ is now
\begin{equation}
\rho_k = \sum_{k'} P_w(k \to k') \rho_{k,k'}
\end{equation}

Solving numerically this equation in the stationary state, it is possible to
determine $\rho^S_k$,
which turns out to be in good agreement with numerical simulations, see
Fig.~\ref{rho_k}. Moreover it is possible to compute the consensus
time $T_N$, which for the voter model turns out to be
\begin{equation}
  T_N=\frac{N \av{k^{1+\theta}}^2}{2\sum_k P(k) k^{2(1+\theta)}
    \rho^S_k(x=1/2)}. 
  \label{eq:20}
\end{equation} 
A remarkable agreement between this expression (evaluated numerically)
and simulations is found
even for some cases where HMF theory fails. Notice that no parameter
is fitted.  Thus, as we can see in
Fig.~\ref{T_HPA}, for small values of $\theta$ ($\theta=0.5)$ and
small $\gamma$, both HMF theory and the HPA provide accurate results
for the consensus time. Larger values of the weight exponent
($\theta=1.5$) are well represented by HPA, while HMF fails. 

 For larger values of $\theta$, however, even the HPA
approximation is not sufficient to capture the correct behavior of the
model.  In this regime, a much harsher
breakdown of the HMF assumptions
occurs~\cite{schneider-mizell09,dynam_in_weigh_networ}: The state of a
node of degree $k$ (or of an edge joining vertices of degree $k$ and
$k'$) depends not only on the degrees but on the detailed
quenched structure of the network, much beyond single-node or
single-pair features. For example, as $\theta \rightarrow \infty$ \cite{dynam_in_weigh_networ}
each node interacts deterministically with its most connected neighbor.
According to HMF equations, which describe an annealed scenario, 
this means that every node will select the most connected node(s) in the network.
However, in a quenched structure each node can choose its partner
only among its neighbors, with the result that different portions of the network
will effectively become independent from the point of view of the dynamics.
Different regions of the network may therefore order in different states, and in this case the 
final global consensus will never be reached (see also \cite{dynam_in_weigh_networ}).


\section{Conclusions}

We have presented a detailed investigation of the
behavior of voter model and Moran processes on weighted complex
networks.  From the analytical point of view we have put forward a
theoretical framework that allows to deal with generic edge
weights. For a specific form of the weights we have derived in detail
all relevant properties of the dynamical processes, such as the exit
probability and the scaling of the consensus time as a function of the
network size.  It turns out that the presence of weights has the
effect of slowing down the Moran process with respect to the
unweighted case, while it generally speeds up ordering with voter
dynamics.  Numerical simulations are in good agreement with
the theory for small absolute values of $\theta$, while for large
$|\theta|$ substantial discrepancies show up. An improved mean-field-like
theoretical approach (heterogeneous pair approximation) taking into
account two-body correlations gives better agreement with
numerics. Still in the limit of large positive (negative) $\theta$,
when the state of a node tends to be deterministically enslaved to the
state of its neighbor with largest (smallest) degree, the theoretical
approaches fail to describe in a satisfactory manner the behavior of
the system.

The positive news is that the mean-field equations describe quite
well the dynamics observed in real (quenched) networks for weight
intensities of the order of the ones observed
in real-world networks \cite{Barrat:2004b}.
However, the generality of this finding, as well as the
intrinsic limits of the theory, are in need of a better understanding
(see also \cite{dynam_in_weigh_networ}).  A theoretical approach able
to take into account the detailed quenched structure of weighted
networks is in order to successfully tackle this problem.

\section*{Acknowledgments}

R.P.-S. and A. B. acknowledge financial support from the Spanish MEC
(FEDER), under project FIS2010-21781-C02-01, and the Junta de
Andaluc\'{i}a, under project No. P09-FQM4682.. R.P.-S. acknowledges
additional support through ICREA Academia, funded by the Generalitat
de Catalunya.  A. B.  acknowledges support of Spanish MCI through the
Juan de la Cierva program funded by the European Social Fund.


\end{document}